 \documentclass[preprint2]{aastex}
 \input{psfig}

\def\gax    {${_>\atop^{\sim}}$}

\def\etal   {{\it et al.}~}

\def\nh     {N$_{\rm H}$~}

\shorttitle{Mathur et al.}
\shortauthors{X-ray Spectrum of a BALQSO}

\begin{document}


\title{Surprises from a deep ASCA spectrum of a Broad Absorption Line Quasar
PHL~5200}


\author{S. Mathur\altaffilmark{1,2},
G. Matt\altaffilmark{3},
P. J. Green\altaffilmark{2},
M. Elvis\altaffilmark{2},
K. P. Singh\altaffilmark{4}
}





\altaffiltext{1}{The Ohio State Univ., smita@astronomy.ohio-state.edu}
\altaffiltext{2}{Harvard Smithsonian Center for Astrophysics}
\altaffiltext{3}{Universit\'{a} Roma Tre, Italy}
\altaffiltext{4}{Tata Institute of Fundamental Research, Mumbai, India}


\begin{abstract}

 We present a deep ($\sim$85~ksec) ASCA observation of the prototype
 Broad Absorption Line Quasar (BALQSO) PHL~5200. This is the best
 X-ray spectrum of a BALQSO yet. We find that (1) the source is not
 intrinsically X-ray weak, (2) the line of sight absorption is very
 strong with \nh$=5\times 10^{23}$ cm$^{-2}$, (3) the absorber does
 not cover the source completely; the covering fraction is $\approx
 90$\%. This is consistent with the large optical polarization
 observed in this source, implying multiple lines of sight. The most
 surprising result of this observation is that (4) the spectrum of
 this BALQSO is {\it not} exactly similar to other radio-quiet
 quasars. The hard X-ray spectrum of PHL~5200 is steep with the
 power-law spectral index $\alpha \approx 1.5$. This is similar to the
 steepest hard X-ray slopes observed so far. At low redshifts, such
 steep slopes are observed in narrow line Seyfert 1 galaxies, believed
 to be accreting at a high Eddington rate. This observation
 strengthens the analogy between BALQSOs and NLS1 galaxies and
 supports the hypothesis that BALQSOs represent an early evolutionary
 state of quasars (Mathur 2000). It is well accepted that the
 orientation to the line of sight determines the appearance of a
 quasar; age seems to play a significant role as well.
\end{abstract}




\section{Introduction}

Broad absorption line quasars, in which the kinetic energy
carried out in the absorbing outflow is a significant fraction of the
bolometric luminosity of the quasar, offer a challenge to our
understanding of the quasar energy budget (as suggested in Mathur, Elvis
\& Wilkes 1995; see also $\S 11.3$ in Krolik 1999). At the same time,
they also offer new insights into the nuclear structure of quasars
(Ogle 1998, Elvis 2001). X-ray observations are important in this investigation
as they offer precise measurements of absorbing column
densities. Combined UV and X-ray analysis is a powerful tool to
understand the physical conditions in the absorbing gas (Mathur \etal
1998 and references there in). However, X-ray observations of BALQSOs
have essentially resulted in non-detections (Green \& Mathur
1996). While the available evidence clearly suggested that the
observed X-ray weakness of BALQSOs is a result of strong X-ray
absorption (Mathur \etal 2000 and references there in), in some
objects absorption was not apparent (Gallagher \etal 1999). So the
nature of X-ray weakness of BALQSOs and the column density of
absorption if any, could not be determined with certainty for the lack
of a good X-ray spectrum.

PHL~5200, the prototype BALQSO at z=1.98 (Burbidge 1968) was clearly
detected in {\it EXOSAT} medium energy experiment (ME), but not by the
low energy experiment (LE) (Singh, Westergaard \& Schnopper 1987). To
obtain consistency between ME and LE requires a column density of \gax
10$^{22}$ cm$^{-2}$ at the source. The detection of PHL~5200 with
EXOSAT made it an excellent candidate for detailed spectral study in
X-rays. A 1994 ASCA observation of PHL~5200 yielded the first X-ray
spectrum of a BALQSO (Mathur, Elvis \& Singh 1995, MES95 here after),
and remained the only one until this year. However, the quality of the
ASCA spectrum was poor and the parameters of the fit were not well
determined in the 17.7 ksec observation. The photon index of the
best-fit power-law was uncertain to $\pm 0.9$ (all the errors are
quoted to 90\% confidence, unless noted otherwise). While excess
absorption with \nh=$1.3^{+2.3}_{-1.1} \times 10^{23}$ cm$^{-2}$ at
the source provided a better fit, a model with only Galactic
absorption was acceptable (see also Gallagher \etal 1999). A good signal to
noise spectrum of a BALQSO
was very much needed and PHL~5200 remained the best target. So we
re-observed PHL~5200 with ASCA, to obtain better quality data and better
constrain the parameters of an X-ray spectrum of a BALQSO. We find that the
spectrum
contains more surprises that affect the interpretation of the BAL
phenomenon.

\section{Observations and Data Analysis}

\subsection{Observations}

 We observed PHL~5200 with ASCA (Tanaka \etal 1994) on 1999 November
 21. ASCA contains two sets of two detectors, SIS (Solid-state
Imaging Spectrometer) and GIS (Gas Imaging Spectrometer). The
effective exposure times in SIS0, SIS1, GIS2 and GIS3 were 83,522
seconds, 73,320 seconds, 93,385 seconds and 93,337 seconds
respectively. The SIS was operated in 1CCD mode with the target in the
standard 1CCD mode position. The GIS was operated in pulse height (PH)
mode. The data were reduced and analyzed using FTOOLS and XSELECT in a
standard manner (see ASCA Data Reduction Guide for
details of data reduction).

\subsection{Data Analysis}

 The source was clearly detected in all the four instruments with
 background subtracted net count rate of 8.6$\pm0.5 \times 10^{-3}$,
 4.8$\pm0.4 \times 10^{-3}$, 5.6$\pm0.6 \times 10^{-3}$, 10.9$\pm0.6
 \times 10^{-3}$ for SIS0, SIS1, GIS2 and GIS3 respectively. We extracted the
 source spectrum from a circular region centered on the source. For
 GIS2 and GIS3 we used the recommended 6$^{\prime}$ radius. For SIS0
 and SIS1 we used 3$^{\prime}$ and 2$^{\prime}$ radii,
 respectively. Larger regions were not possible as the source was
 quite close to the chip gap. Background events were extracted from
 the the identical regions for each instruments from the deep
 background images supplied by the ASCA guest observed facility. For
 GIS, the latest (March 1995) response matrices were used, while for
 SIS, they were generated using FTOOL {\bf sisrmg}. The spectra were
 then corrected for vignetting and the channels were grouped to
 contain at least 15 net counts per channel. In summary, the spectra
 for all instruments were extracted in a standard manner.

 We also extracted background from different source-free regions on
 the same detector from our PHL~5200 observation. We found that the
 spectra extracted in this way did not alter our results in any
 significant way. For the rest of the paper we will discuss only the
 spectra with background extracted from the deep background
 observations.

 Our observations were taken towards the end of ASCA mission, and the
 CCDs on the satellite were significantly degraded over time. There
 are problems with the charge transfer inefficiency (CTI), the
 residual dark distribution (RDD), and an unidentified additional loss
 of SIS low energy efficiency.\footnote{For ASCA Calibration
 Uncertainties see:
 http://legacy.gsfc.nasa.gov/docs/asca/cal\_probs.html} For our
 observation in the 1CCD mode and for a weak source like PHL~5200, the
 latter is the most serious issue. The loss of the low energy response
 of the two SIS detectors can be parameterized with excess absorption
 (Yaqoob \etal 2000). This ``\nh correction'' is different for SIS0
 and SIS1 and is a function of time. We have taken into account all
 the calibration uncertainties carefully in our spectral extraction
 and analysis.

\subsection{Spectral Analysis}

 The extracted spectra were  analyzed
 using {\bf XSPEC} version 11.0 (Arnaud, 1996). Data below 0.6 keV for
 SIS, below 0.9 keV for GIS and above 9.5 keV for all the instruments
 were ignored in spectral fits as these energy channels are not well
 calibrated. We first fitted the spectra from each instrument
 separately and found the results to be generally consistent. We then
 fitted all the spectra simultaneously leaving the normalizations
 free, to allow for small differences in absolute calibration.

 As a first step we fitted the spectra with a simple power-law and
 Galactic absorption (with \nh=$4.8 \times 10^{20}$ cm$^{-2}$, Stark
 \etal 1992). The ``\nh correction'' discussed in section 2 was
 applied to the SIS spectra, which for our epoch of observation was
 $6.8 \times 10^{20}$ cm$^{-2}$ for SIS0 and $9.8 \times 10^{20}$
 cm$^{-2}$ for SIS1, with $\approx \pm 10\%$ uncertainty. The fit was
 not good, with $\chi^2=226.5$ for 137 degrees of freedom. As excess
 absorption was indicated in the earlier spectrum of PHL~5200 (MES95)
 and in other BALQSOs (Mathur \etal 2000 and references there in),
 absorption at the source was added as a next step. The fit was
 better, but clearly showed negative residuals around one keV, and
 positive residuals at lower energies (Fig. 1, Fig. 2). Note that the
 behavior of the low energy residuals is exactly opposite to that due
 to the low energy calibration problem which produces spurious
 negative residuals. The implied recovery of the
 underlying power-law is a typical signature of partial covering by
 the absorber.  So we fitted the spectrum with a partial covering
 model and the fit was significantly better ($\chi^2=182.5$, 135
 degrees of freedom). The resulting \nh=$5\pm1 \times 10^{23}$
 cm$^{-2}$ and covering factor is $0.9^{+0.05}_{-0.06}$. The
 confidence contours of the interesting parameters are plotted in
 figure 3. The scale over which figure 3 is plotted shows the range of
 $3\sigma$ uncertainty in the earlier observation (MES95). The column
 density at the source is greater than $2\times 10^{23}$ cm$^{-2}$ and
 the power-law slope $\alpha$ is greater than 1.2 at 99\% confidence (flux
$f(E) \propto E^{-\alpha}$ where E is the energy).

The only residuals left after the partial covering model fit, are
those above $\sim 8$ keV (observed frame), lying below the model
values (Figure 2).  Such residuals are not expected to be due to any
calibration problems. They imply that a simple power-law continuum
must turn over at higher energies. So we added a high energy cutoff to
the model and re-fitted the data. This resulted in a much better fit
with $\chi_{\nu}^{2}= 1.18$. This ``best fit'' model, however,
resulted in an unusually low value of the high energy cut-off of
E$_{cuoff}=18$ keV (rest-frame) with e-folding energy of 0.4 eV. The
typical high energy cutoff in quasar spectra is believed to be about
300 keV, though Yaqoob (2000) has drawn attention to the fact that
this value is not well measured. We are aware of at least one another
AGN with a very low cutoff value. In ESO 103-G35, Wilkes \etal (2001)
found E$_{cutoff}=29\pm 10$ keV. Low values of E$_{cuoff}$ are likely
to be important in understanding the spectrum of the cosmic X-ray
background (Yaqoob, 2000). We note, however, that the negative
residuals observed in our spectrum of PHL~5200 are significant only in
the last few channels. So we caution against attaching too much
significance to this result.

 As discussed above, the source is highly absorbed with \nh=$5\times
 10^{23}$ cm$^{-2}$. While the absorber is not Thomson thick, the
 effects of Thomson and Compton scattering may start becoming
 important at such column densities. In the analysis above, only
 photoelectric absorption is taken into account. The negative
 residuals observed at higher energies could be the effect of excess
 absorption due to Compton scattering, though for the observed column
 density it is not expected to be a significant effect (Matt \etal
 1999). Nevertheless, we re-fitted our spectra with a partial covering
 model that fully incorporates Compton scattering in the absorber, as
 discussed in Matt \etal (1999). The resulting parameters are: \nh=$6\pm 2
 \times 10^{23}$ cm$^{-2}$, $\alpha=1.9\pm0.4$, and covering fraction
 =$0.94\pm 0.04$. These are consistent with the model without
 Compton scattering, as expected.

For a spherical distribution of matter with \nh=$5 \times 10^{23}$
 cm$^{-2}$, a Fe-K emission line with equivalent width (EW) of 270 eV
 is expected (as measured against the observed continuum). We do not
 detect any line at the position of Fe-K$\alpha$ (rest energy 6.4
 keV). The upper limit on the observed EW for a narrow line is 71 eV
 for SIS0 data and 102 eV for GIS3 data. The corresponding rest-frame
 EWs are 212 eV and 304 eV respectively. So we cannot put any
 meaningful constraint on the geometry of the absorber as viewed by
 the continuum source.

\section{Discussion}

 The BALQSO PHL~5200 is clearly detected in all the instruments in
 this deep ASCA observation. With a total of 2610 ``good'' counts,
 this is the best X-ray spectrum of a BALQSO yet. The parameters of
 the spectral fit are well constrained; the X-ray flux is
 highly absorbed with N$_{\rm H}=5\pm 1 \times 10^{23}$
 cm$^{-2}$. While consistent with the conclusions of MES95,
 the earlier low signal to noise spectrum was also consistent with a
 model with no excess absorption (see also Gallagher \etal 1999). We
 unambiguously confirm large absorbing column density in PHL~5200. The
 X-ray weakness of BALQSOs thus is surely a result of absorption.

 The absorber does not cover the continuum fully, with a
 covering fraction 90$\pm 5$\%.  This must be the reason for the
 apparent lack of excess absorption and flatter spectrum in the low
 signal to noise spectrum (see also Green \etal 2001). Partial
 covering of the absorber is consistent with the large optical
 polarization observed in this source (Goodrich \& Miller 1995, Cohen
 \etal 1995) implying that there are at least two lines of sight
 to the nucleus: one direct, highly absorbed, and another scattered
 but unabsorbed. Gallagher \etal (1999) have discussed the possibility
 that partial covering by the absorber might be responsible for highly
 polarized BALQSOs to be relatively X-ray bright. Here we find that indeed,
 that is the best fit model to the PHL~5200 spectrum.

 An associated absorber with high column density and partially
 covering the continuum is our preferred interpretation of the
 observed spectrum of PHL~5200. While physically motivated, this is
 not the only way to model the ASCA spectrum. For example, the low
 energy recovery of the absorbed spectrum may be be modeled as an
 unrelated soft-excess.\footnote{The nearby object, $\approx
 8^{\prime\prime}$ away from the optical position of PHL~5200, is an
 early K star with no emission lines, and is therefore very unlikely
 to contribute to the X-ray flux of PHL~5200.} The dip at $\approx 1$
 keV is also a signature of a warm absorber. Note, however, that in
 PHL~5200 it corresponds to rest-frame energy of $\approx 3$
 keV. Since absorption due to highly ionized sulphur or silicon occurs
 at these energies, such an interpretation might be
 plausible. However, it seems unlikely given that these are not among
 the most abundant elements.

 The most surprising result of this observation is the steep spectrum,
 with X-ray power-law slope $\alpha =1.7\pm 0.4$. This might be an
 additional reason behind the unexpected non-detections of BALQSOs
 even in sensitive hard X-ray observations ($\S 1$).  The mean ASCA
 slope for radio-quiet quasars at high redshifts is $\alpha=0.67\pm
 0.11$, with a dispersion of $\sigma=0.07$ (Vignali \etal 1999, in the
 reshift interval between z=1.9 and 2.3).  Our observations imply that
 the spectra of BALQSOs may {\it not} be exactly like other
 radio-quiet quasars, but are steeper.  The observed, absorption
 corrected, 2--10 keV flux is $3.7 \times 10^{-13}$ ergs cm$^{-2}$
 s$^{-1}$ (SIS0), consistent with the 1994 ASCA observation
 (MES95). So, the observed steep spectrum is unlikely to be a result
 of a short lived high-state. The only other BALQSO, PG 2112+059, for
 which a spectrum is available, has $\alpha=0.98^{+0.4}_{-0.27}$
 (Gallagher \etal 2001). As noted by Gallagher \etal, this is
 consistent with the mean quasar slope. However, it is also consistent
 with as steep as 1.38 at 90\% confidence.

 As discussed in section 2.2, CCDs on ASCA have been degraded over
 time. While we have been very careful in our analysis, there might be
 some unknown calibration uncertainties which have affected the
 results presented here. Future observations with Chandra and
 XMM-Newton will be useful in this respect. The discussion below assumes
 that these ASCA results are correct.

 At low redshifts, Brandt, Mathur \& Elvis (1997) found that the hard
 X-ray spectra of Seyfert galaxies are typically flatter than about
 $\alpha=1.0$. The Seyfert galaxies with steeper spectra are the
 narrow line Seyfert 1 galaxies (NLS1s).  Brandt (2000) has discussed
 the analogy between low ionization BALQSOs and NLS1 galaxies. Mathur
 (2000) has discussed the analogy between BALQSOs and NLS1s further
 and has argued that NLS1 may be active galactic nuclei (AGNs) in the
 making. In this scenario, young radio-quiet AGNs are accreting at a
 high Eddington rate and have steep spectra. Over time, the accretion
 rate drops and the X-ray spectrum flattens. BALQSOs may be in that
 early evolutionary phase when the shroud surrounding the nuclear
 black-hole is being blown away and a quasar emerges (e.g. Fabian
 1999, see also Becker \etal 2000). If the observed steep spectrum of
 PHL~5200 is a general property of BALQSOs, then it supports the
 evolutionary hypothesis of Mathur (2000) and further supports their
 analogy with NLS1 galaxies. On the other hand, PHL~5200 is one of the
 X-ray brightest BALQSOs, and so may not be representative of the
 BALQSO population. What is more likely, however, is that its
 orientation and cloud geometry fortuitously allow for substantial
 reflection of X-rays into our line of sight.

 It has been an accepted paradigm that the orientation to the line of
 sight determines the appearance of an AGN. In addition to
 orientation, the age of a quasar also seems to play a role in its
 appearance. The steep X-ray spectra in NLS1s are a result of their
 high accretion rate, close to the Eddington limit (Pounds \etal
 1995). BALQSOs, however, are much more luminous than NLS1s, and so
 must have far more massive black holes. Steep X-ray spectra in such
 systems may prove to be challenging to accretion theories.





\acknowledgments We thank Tahir Yaqoob for help with ASCA calibration
uncertainties, and Perry Berlind for obtaining the optical spectrum of
the K star at the Fred Lawrence Whipple Observatory.  This work is
supported in part by NASA grants NAG5-9270 \& NAG5-8913 (LTSA) to SM.

\clearpage
\thispagestyle{empty}


\newpage
\begin{table}[h]
\caption{Spectral fits to ASCA data of PHL5200}
\begin{tabular}{|lcccc|}
\tableline
Model& $\alpha_E$ & N$_H$ (free)$^a$&
 Additional Parameter & $\chi^2$ (dof)$^c$\\
&&&&\\
\tableline\tableline
\multicolumn{4}{|l}{Power-law +N$_H$ (Gal.) fixed:} & \\
 & 0.5$\pm 0.1$ & & & 226.5 (137)\\
\multicolumn{4}{|l}{Power-law +N$_H$ (Gal.) fixed +N$_H$ (z=1.98):} & \\
 &0.9$^{+0.2}_{-0.3}$& 8$^{+4}_{-5}$ && 212.5 (136)\\
\multicolumn{4}{|l}{Power-law + Partially Covering Absorber} & \\
 & 1.78$\pm 0.4$ & 50$\pm 14$ & Covering fraction= 0.90$^{+0.05}_{-0.06}$ &
182.5 (135)\\
\multicolumn{4}{|l}{Power-law + Partially Covering Absorber, Including Compton
Scattering} & \\
 & 1.9$\pm 0.4$ & 62$^{+19}_{-17}$ & Covering fraction= 0.94$^{+0.03}_{-0.06}$
& 162.8 (132)\\
\multicolumn{4}{|l}{Power-law + Partially Covering Absorber + High Energy
Cutoff} & \\
 &1.4$\pm 0.4$& 46$^{+16}_{-17}$ & Covering fraction= 0.85$^{+0.07}_{-0.14}$ &
157.3 (133)\\
&&&&\\
 &&                          & Cutoff Energy= 18.0$^{+0.4}_{-1.0}$ & \\
\tableline
\tableline
\end{tabular}
\smallskip
a: $\times 10^{22}$ cm$^{-2}$\\
b: in units of $10^{-4}$ photons keV$^{-1}$ cm$^{-2}$ s$^{-1}$ at
1 keV\\
c: degrees of freedom.
\end{table}



\clearpage

\figcaption{Residuals ($\Delta \chi^2$) to a power-law fit showing
clear signature of a partially covering absorber. Here the power-law
was fitted only to the data above 2 keV and extrapolated to lower
energies. Note the strong turn-over below 2 keV and recovery below
about 1 keV. Only SIS data are shown for clarity; dots: SIS0, triangles: SIS1.}

\figcaption{ASCA data of PHL~5200 divided by fitted models: (from top
to bottom) (a) power-law, Galactic absorption, plus excess absorption
at the source. Note the strong recovery at low energies.; (b) a
partial covering model. The residuals left are at energies \gax 8 keV;
(c) additional high energy cutoff. Dots: SIS0, Circles: SIS1, Triangles: GIS2,
Filled triangles: GIS3}

\figcaption{Confidence contours of the power-law photon index against
the absorbing column density (the photon index is $=1+\alpha$). Note
the large best fit column density and the steep power-law
spectrum. The horizontal lines represent the average spectral slope
and range in z\gax 2 radio-quiet quasars. The scale over which the
plot is made shows 3$\sigma$ uncertainty in the previous
observation. }

\newpage
\clearpage
\begin{figure} [h]
\psfig{file=fig1.ps,height=2.in,width=6.0in,angle=-90}
\end{figure}
\clearpage

\newpage
\clearpage
\begin{figure} [h]
\psfig{file=fig2a.ps,height=2.in,width=6.0in,angle=-90}
\psfig{file=fig2b.ps,height=2.in,width=6.0in,angle=-90}
\psfig{file=fig2c.ps,height=2.in,width=6.0in,angle=-90}
\end{figure}
\clearpage

\newpage
\clearpage
\begin{figure} [h]
\psfig{file=fig3.ps,height=3.0in,width=5.0in,angle=-90}
\end{figure}
\clearpage


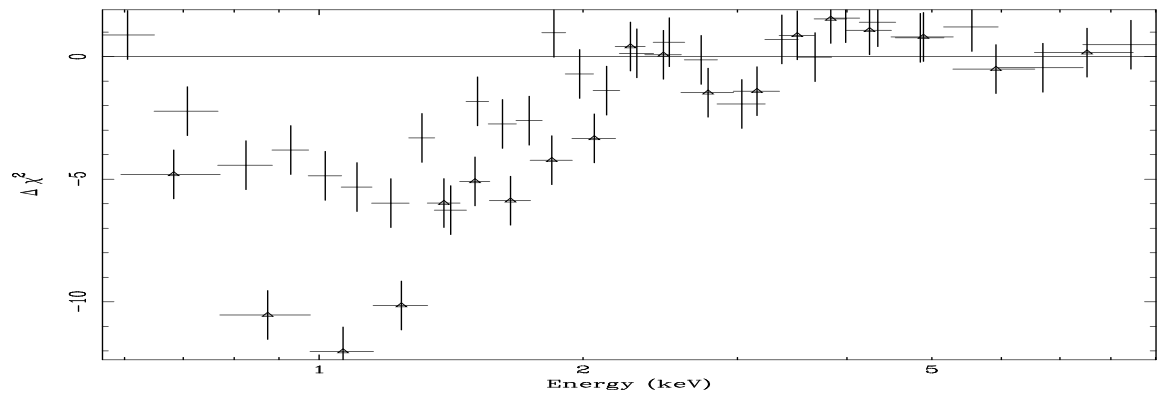
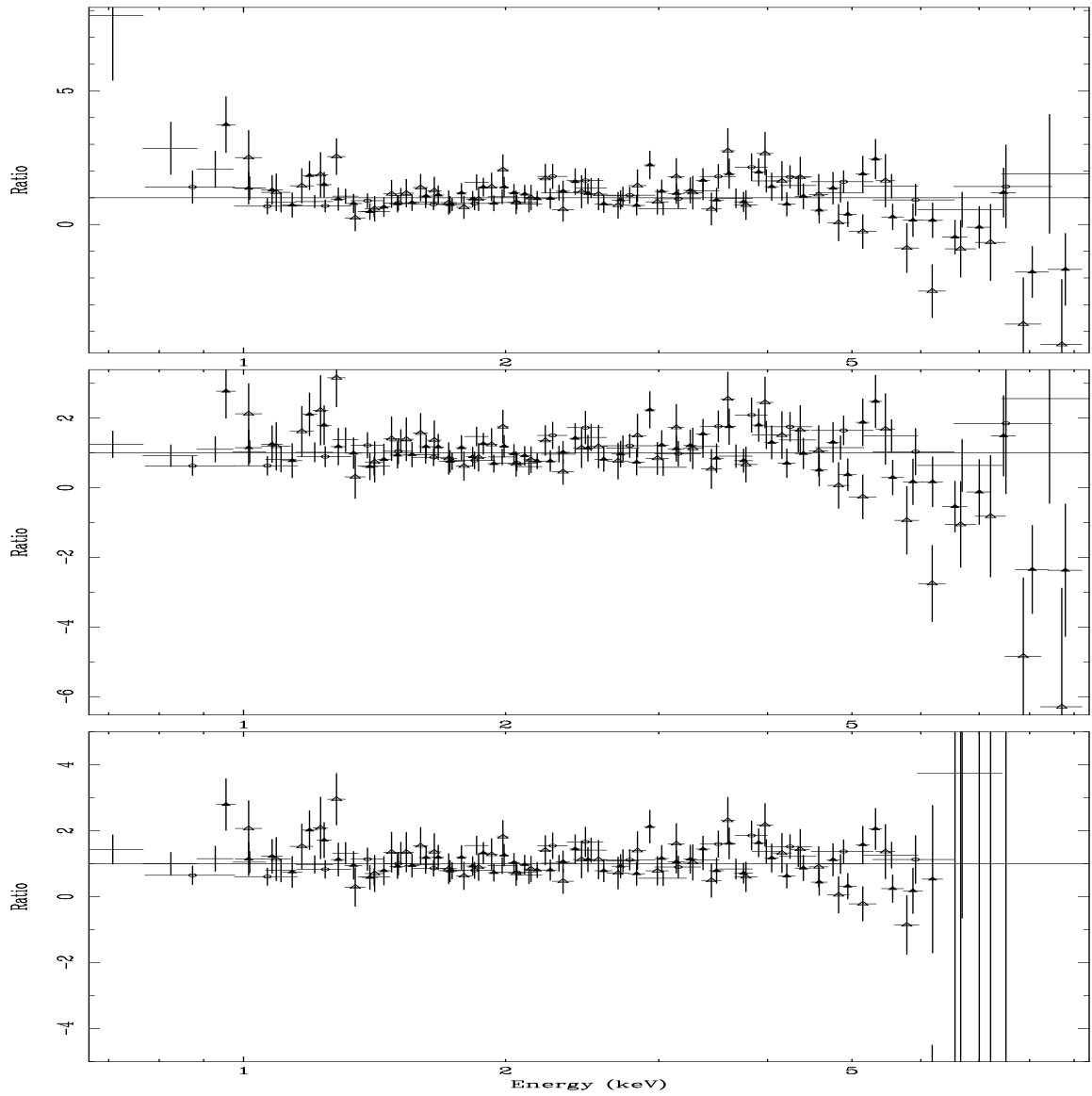
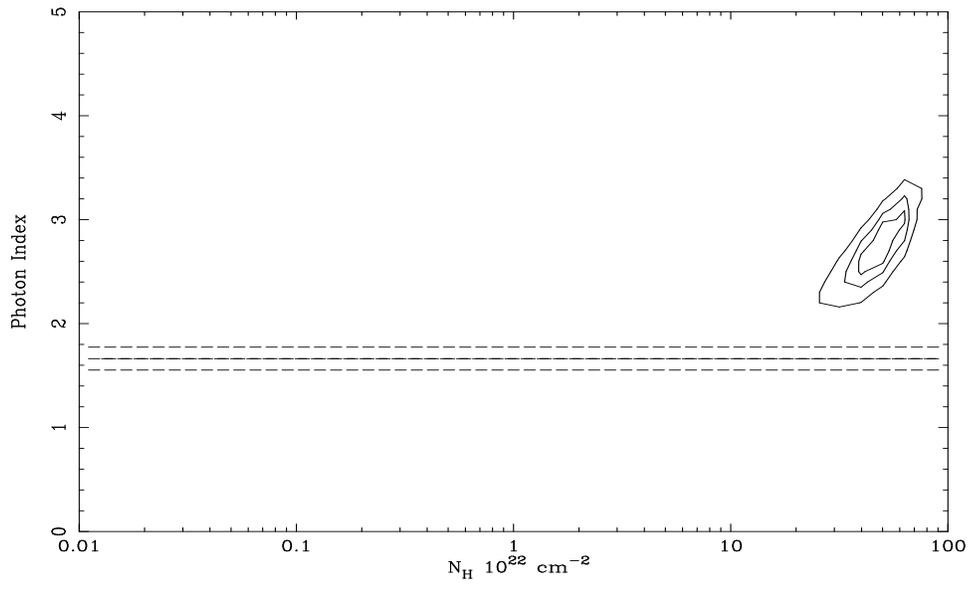
\end{document}